# Local Interactions and the Emergence of a Twitter Small-World Network

Eugene Ch'ng
*School of Computer Science*
*Big Data and Visual Analytics Lab*
*University of Nottingham Ningbo China*
*Science and Engineering Building*
*199 Taikang East Road, 315100 Zhejiang Ningbo, China*
eugene.chng@nottingham.edu.cn

**Abstract**

The small-world phenomenon is found in many self-organising systems. Systems configured in small-world networks spread information more easily than in random or regular lattice-type networks. Whilst it is a known fact that small-world networks have short average path length and high clustering coefficient in self-organising systems, the ego centralities that maintain the cohesiveness of small-world network have not been formally defined. Here we show that instantaneous events such as the release of news items via Twitter, coupled with active community arguments related to the news item form a particular type of small-world network. Analysis of the centralities in the network reveals that community arguments maintain the small-world network whilst actively maintaining the cohesiveness and boundary of the group. The results demonstrate how an active Twitter community unconsciously forms a small-world network whilst interacting locally with a bordering community. Over time, such local interactions brought about the global emergence of the small-world network, connecting media channels with human activities. Understanding the small-world phenomenon in relation to online social or civic movement is important, as evident in the spate of online activists that tipped the power of governments for the better or worst in recent times. The support, or removal of high centrality nodes in such networks has important ramifications in the self-expression of society and civic discourses. The presentation in this article anticipates further exploration of man-made self-organising systems where a larger cluster of ad-hoc and active community maintains the overall cohesiveness of the network.

## 1. Introduction

Small-world networks are classes of real-world networks which interpolate between random networks [1] and regular lattice [2], a characteristic of complex systems, which exist in a state between order and disorder [3,4]. Small-world networks are also characterised by average shortest path length and high clustering coefficient [5,6], and is associated with efficient information transfer, global integration and regional specialisation. Man-made, physical and biological systems that are self-organising have a small world characteristic, these are neural networks [2], food webs [7], computer networks [8], social networks [9], and bilateral trade [10].

Whilst there has been no lack of theoretical and experimental studies related to small-world networks since it was first discovered, there appears to be no formal description of the dynamical development of a certain class of the real-world small-world phenomenon where two opposing communities, in the course of arguments (local positive and negative interactions) over certain ideology, initiated the global emergence of a small-world network, and maintained the cohesiveness of it.

The interest of this study lies in the controversial online #FreeJahar activities calling for the freedom of the Boston bombing suspect Dzhokhar Tsarnaev only because teenage girls believed he is "too beautiful to be a terrorist" [11]. The younger Boston suspect have since become a teen heartthrob as thousands of girls express their love for the bomber in worrying online forums [12]. Facebook with 8,681 members at present (from 6,500 members on 12 May 2013), in the "Dzhokhar Tsarnaev Free Jahar" group and Tumblr tribute accounts were set up in support of the teen with the #FreeJahar Twitter tag, which were trending during the days when the activities first appeared. Whilst the #FreeJahar group and its tweets were controversial, online fan clubs and "killer crushes" for suspected killers are not new [13]. But, as noted on The Verge [14], for every #FreeJahar tweet, there is a user expressing vitriolic disgust that Tsarnaev supporters exist – "#freejahar What!? He's a #Terrorist scumbag and he deserves to fry," and "all u sympathizers can burn in Hell too." This opposing force seems to be a catalyst for the formation and maintenance of the small-world network.

Recent world events have indicated that civic liberty have found important channels in which to express themselves, one of the most important of which is social media, and Twitter seems to be a popular choice. On the surface, Twitter appears to be a micro-blog and a short messaging service that supports only 140 characters in each tweet, presented on users' displays in chronological order. Twitter is also a social network due to the provision of a follower-followee service. The follower-followee network however, does not truly represent an active network. A deeper, more genuine representation of an active network can be mined when streamed Twitter activities are mapped in a certain way (see subsequent sections). Activity networks can reveal a truer picture of online interactions between activists. Such online activities, and coordination can potentially translate into physical uprising, such as the Syrian civil movement [15,16]. This is possible as the span of time and space is reduced via social medium as catalysts. Understanding the small-world phenomenon in relation to online social or civic movement is important, for the support, or removal of high centrality nodes in such networks has important ramifications in the self-expression of society and civic discourses.

## 2. Methods: Mapping Genuine Twitter Activities

The dataset investigated here consists of 5 hours of Twitter activity related to the #FreeJahar movement. The particular dataset was selected from 210 datasets collected over 45 days, as it was the largest dataset showing trending news and heightened interaction activities. A Big Data architecture [17] was used for collecting and collating the dataset via the Twitter API using the keywords #Dzhokhar, #FreeJihad, #FreeJahar, #Tsarnaev. The network is multimodal and consists of two types of nodes within the network, the first is a user node (a Twitter account), and the second is a tweet node which has possible mentions of other users. If A tweeted B and C, the graph would be A→TweetFromA→B and A→TweetFromA→C, with A as a single node with two directed targets. It is possible that B will reply to A, in which case the graph would link from B→TweetFromB→A.

The dataset selected for this study ranked highest in terms of the number of nodes and edges, retweets, clustering coefficient [5,6], modularity [18,19], number of shortest path, and shortest average path length, making it a particularly interesting dataset to investigate. Figure 1 is the growth and development of the small-world network. The next section describes the activities that led to the formation of a small-world network.

Private account names are not shown, but an associated #FreeJahar symbol is used throughout in the cluster. They contain related names ('jahar', 'tsarnaev', 'troy', 'crossley', 'dzhokhar', etc), mutations of the names ('crussley', 'tsar', etc), and verbs and adjective concatenated with the names ('free', 'let_him_go', 'innocent', 'family', etc). Opposition community has names associated with keywords such as 'faithful', 'usa', 'bible', etc.

## 3. Results

This section demonstrates the result of the mapping technique in the previous section.

### 3.1.     The Formation of a Small-World Network

The growth of the network, and how information is transferred, can be witnessed in the time-based records in Figure 1. At 2.15pm, the network density is particularly sparse, with disjointed nodes scattered across the social information landscape. Conversations related to the #FreeJahar social movement expressed as a cluster of connected nodes located at the top appeared early in the graph. Larger circular clusters of retweets directly below the social movement belonged to major news channels. These are @CNN, @BreakingNews, @NBCNews, @Guardian, @YahooNews, @BBCBreaking, @BostonDotCom, @[Private Account, Boston], @BostonGlobe, as well as associated news correspondents. @BBCBreaking is connected to @YahooNews to the rest of the network, and the @Guardian is connected via @CNN, @NYTimes, @ggreenwald, a private account, and the social movement to the rest of the network.

At 2.30pm, several Twitter users separately picked up a news item related to the event "FBI kills Orlando man with links to the Boston Bombers during interview" (paraphrased to include similar tweets across the network). As information diffuses from news sources and Twitter users picked them up and forward them to their follower networks, the nodes begin to connect. The density of the network increases as retweets, conversations and links are connected into a cohesive whole, forming a small-world between major news channels, and Twitter users within the #FreeJahar movement and their opposers.

Disconnected communities carrying news items that are not associated with news channels are also present, such as the large circle at the bottom of the figure. This belonged to a mass of followers following a single Twitter user @[Private account, Boston] in a socially isolated community of a topical interest. Clusters of smaller networks at the bottom of the figure are news reaction of the same topic, but in different languages. Within the same area, other smaller disconnected network clusters are of a different tweet type, they relate to "slain Tsarnaev friend confessed to Waltham triple murder" (paraphrased). Isolated dyads and triads scattered in the landscape are related but separate retweets and conversations between friends that are not of the #FreeJahar movement. These were later removed (Figure 2) for graphical clarity

As seen in the graph, the small-world network is mainly connected via private accounts and news correspondents. One of the most active connections to news channels came from the #FreeJahar cluster. The first connection to the cluster is @CNN, followed by @BostonDotCom, @Guardian, @YahooNews, @ggreenwald (Guardian correspondent), and @BostonGlobe. The first-order zone is later connected by the second-order zones.

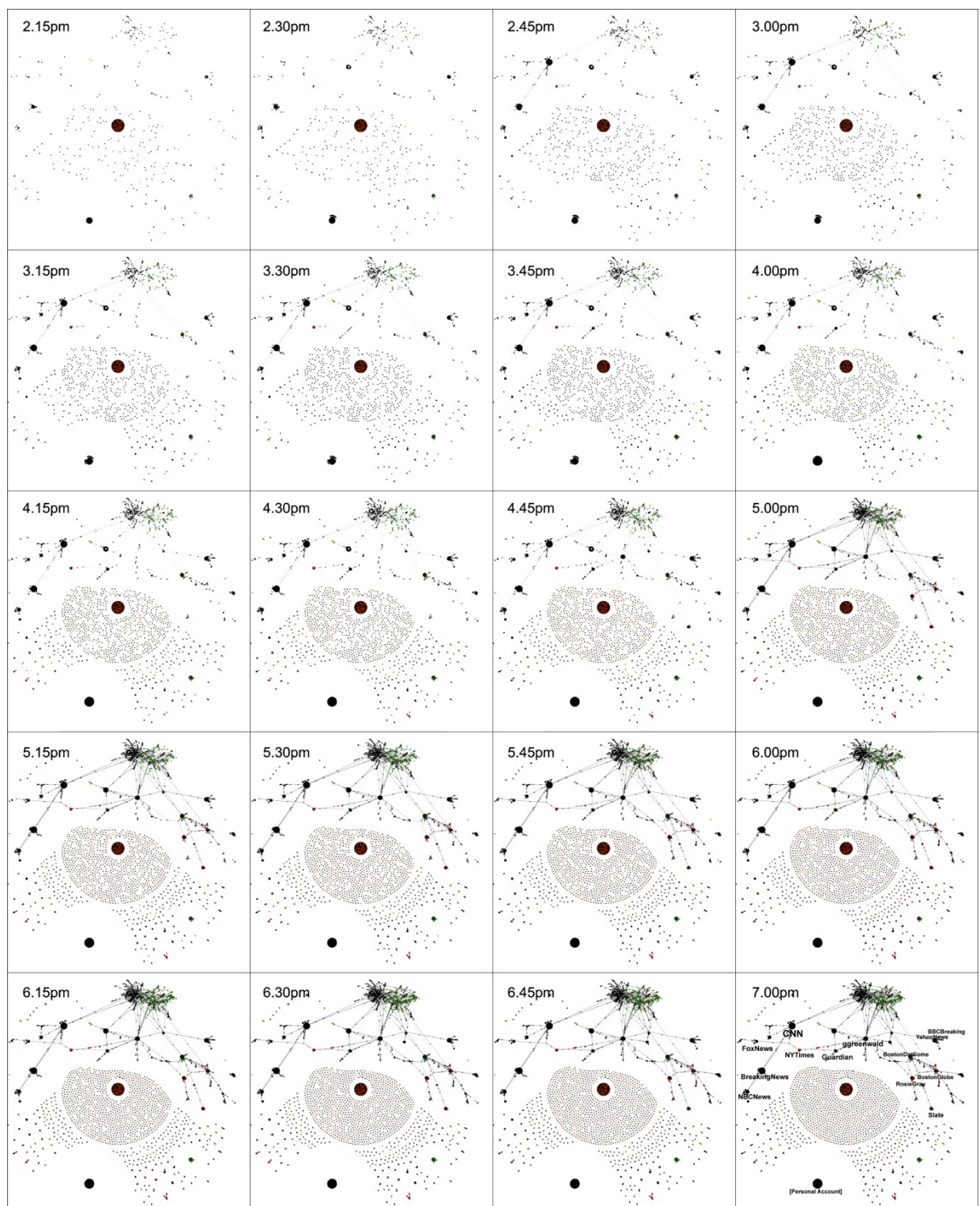

*Figure 1. A dynamic social information landscape showing the growth and connection of the dataset forming a small-world network as news went viral and communities and the #FreeJahar social movement pick up news related to the FBI killing of a suspect.*

### 3.2. The Maintenance of the Small-World Network

This section delves into the communities and structural properties of the isolated network that maintains the small-world phenomenon. Figure 2 is a visual reconfiguration of the same dataset without any changes to the structure of the network. A difference can now be seen between retweets and activities. Tweets produced by the reactions of diverse news sources have circular expressions and community activities are clustered but disorderly.

The major news event relates to the topic 'man shot and killed by FBI in Florida knew bombing suspect Tamerlan Tsarnaev says friend'. The figure shows distinct expression of retweets and activities composed of conversations, the source of the conversation and retweets (black nodes). The graph on the left shows larger nodes that ranked highest in the degree centrality. The red coloured nodes have high degree centrality, these are larger nodes orbited by retweet nodes and their users (as non-black coloured nodes). The graph on the right shows larger nodes that ranked higher in Betweenness centrality, redder colour nodes ranked higher in Closeness centrality. The nodes which have higher Closeness centrality are, on average, closer to all other nodes. Nodes ranked higher in Betweenness centrality are the bridge between communities, they are the gatekeepers of information.

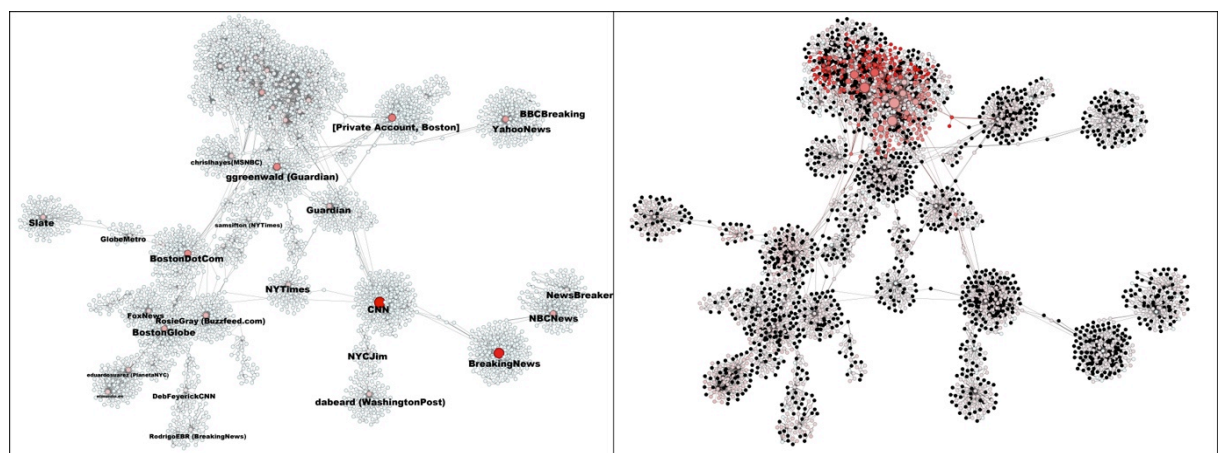

*Figure 2. Two versions of the small-world network. The graph on the left shows the degree centralities, larger redder nodes have more retweets. The graph on the right shows the Betweenness centrality, the larger redder cluster of conversations belonged to the #FreeJahar community, indicating that the cluster is central in the small-world, which maintains the network with heightened activities. Black nodes are retweets*

Figure 3 (top left) is a close-up of the #FreeJahar cluster consisting of supporters and opposers (transparent green patch), clearly separated into two opposing but cohesive communities. Opposer nodes closest to the '#FreeJahar Clusters' have more intense interactions with the #FreeJahar group. The semantics range from statements, quotes and retweets from news and intense emotional arguments between the two groups. Most arguments were criticisms. The special configuration of nodes and edges appears to divide the two opposing groups into very distinct boundaries. Within the transparent green patch, the opposer group furthest away from the edge of the patch have the least interactions with the #FreeJahar cluster. The two adjacent clusters together are loosely organised. Even though some nodes are highly active and with important roles within the community, there appears to be no leader.

Observation of the evolution of dataset over time (Figure 1 and 4) points to factors that connect isolated communities into a cohesive network within a timescale of 5 hours. A small-world network typically has a clustering coefficient greater than a random network (CC≈0) and an average path length that is greater than a random network (PL≈3) and much lesser than a lattice [6]. At 2.15pm, the average clustering coefficient and average path length were the lowest (*CC=0.153, PL=1.069*), and communities were largely disconnected. As time progresses, at 2.30-2.45pm, the #FreeJahar cluster became more active, responding to a news event, which connects the cluster with major news channels (@CNN, @BostonDotCom by #FreeJahar group and opposers). At 4.15pm (*CC=0.272, PL=6.534*), the first-order zone, e.g., between YahooNews and ggreenwald of the Guardian and the social movement is bridged by the community, located at the top right of the cluster. At 4.45pm, the first-order zone (the news channels) connects second-order and tertiary zones. From 5pm (*CC=0.292, PL=6.767*) onwards, the small-world network is reinforced as more links are bridged, consummating in a small-world network at 7pm with high average clustering coefficient and path length (*CC=0.298, PL=7.276*). It will take an average of 7 steps for information, via intermediate nodes, to reach other communities. The analysis shows that the #FreeJahar community is the contributing factor that connects the disparate community into a small-world network; the community also maintains the ties via the strengthening of the interactions between the communities. This is a natural reaction as news items are used for accusations and arguments. As figure 1 suggests, it pulls in isolated clusters of community, without which the small-world phenomenon is not perfect. It is unlikely that the members of the community are aware that the small-world network originated from their activities. The members are bounded by their self-directed goals, without realising that a global structure emerges from their local interaction.

The graph shows that information is bidirectional – #FreeJahar activities connect the small-world whilst news events were used as arguments and criticisms, which strengthens and maintains the network.

News channels ranked highest in the degree centrality, whereas #FreeJahar cluster ranked highest in the Betweenness and Closeness centrality measures. This is another strong indication that the small-world network between the communities is maintained and strengthened by community activities. The centrality of the cluster corresponds to Leavitt's observation [20], that "where high centrality, and hence independence are evenly distributed, there will be no leader, many errors, high activity, slow organisation, and high satisfaction". The edges that play a central role in connecting the small-world network can be traced within 2 steps of nodes with high Betweenness centrality within the community, these highly active nodes are important in the maintenance of the network. Removing these nodes will certainly disrupt the network. Moody and White [21] observe, "A group is structurally cohesive to the extent that multiple independent relational paths among all pairs of members hold it together". In this context, removing the agents that keeps the cluster alive will invariably disrupt the community. Moody and White's structural cohesion, defined as "the minimum number of actors who, if removed from a group, would disconnect the group" occurred when many of the accounts belonging to the supporter group have been suspended or deactivated, which resulted in fractures in the network in latter datasets (Figure 3).

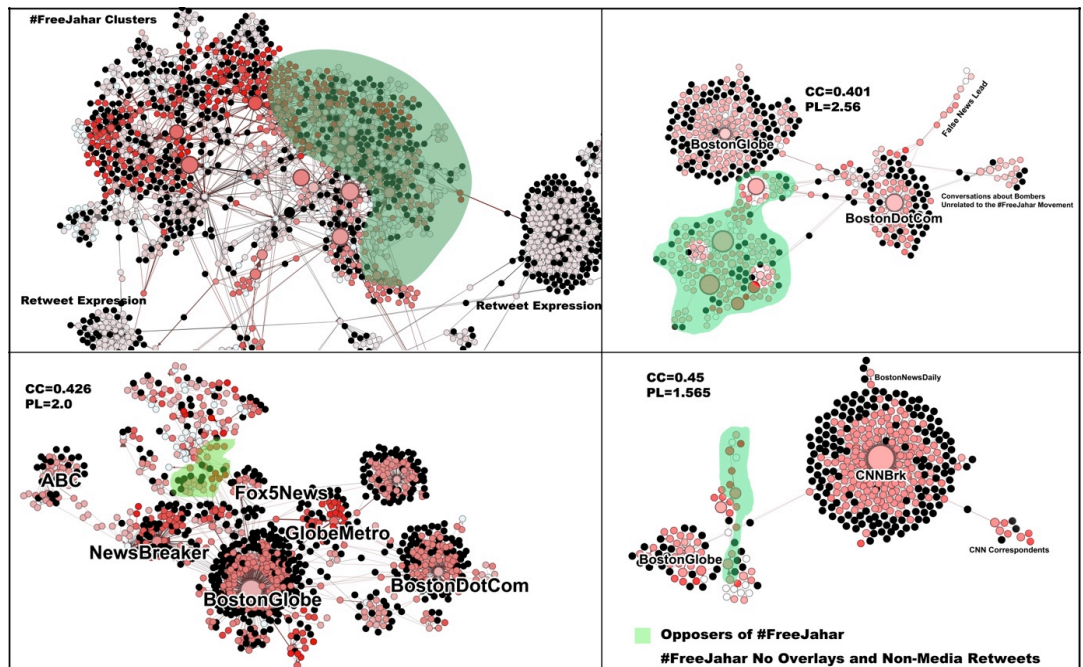

*Figure 3. The fracturing of the network due to important nodes with high Betweenness centrality being removed from Twitter*

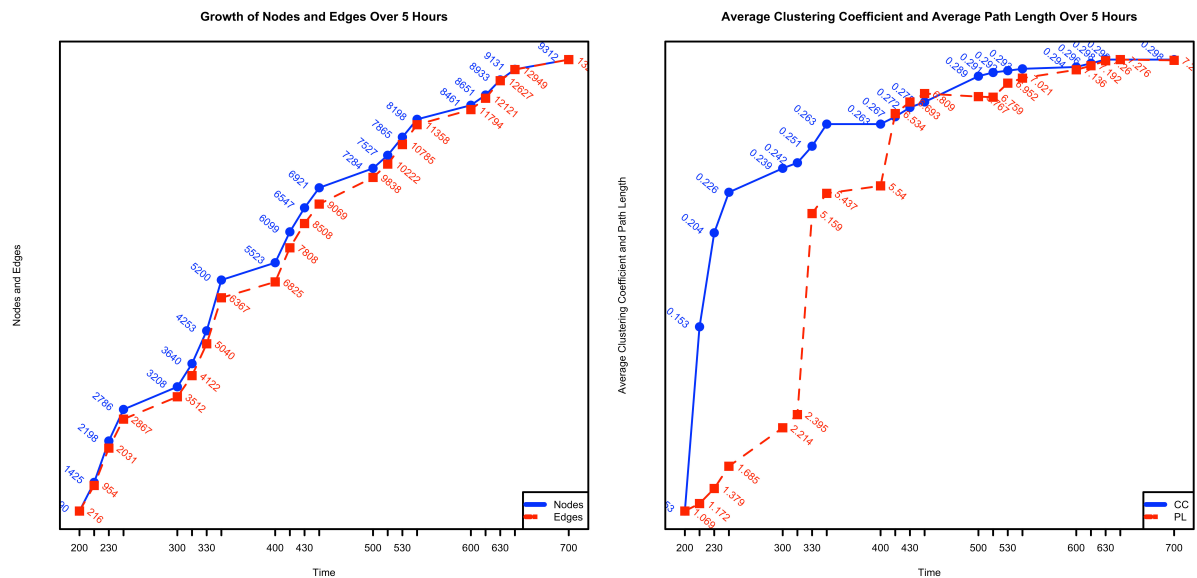

*Figure 4. Graphical plots showing (left) the node-edge growth and (right) the clustering coefficient and average path length over time*

## 4. Conclusion

Efficient information diffusion is important in coordination efforts in both offline and online systems. It is a well-known fact that small-world networks are highly efficient as compared to lattice or random networks. Small-world networks have global integration of clusters and local specialisation. In the context of this study, local specialisation can be seen in the community clusters and news sources with retweets, and global integration is the connection of each of these clusters into a small-world network. The nature of a small-world network is a catalyst for efficient information diffusion as it takes an average of 6 steps in order to reach another unknown individual, as experiments demonstrate [22,23]. Kochen [24] suggested that "it is practically certain that any two individuals can contact one another by means of at least two intermediaries. In a [socially] structured population it is less likely but still seems probable. And perhaps for the whole world's population, probably only one more bridging individual should be needed." Actual online activities facilitated by social media are no different.

The mapping of online social movement within the limited Twitter interface and service reveals that instantaneous events and connections between individuals and communities can result in a small-world network. A close inspection reveals that these small-world networks have a direct relationship with active users and clusters such as the #FreeJahar movement composed of supporters and opposers, which exists solely as a reaction to the controversial movement. As seen in the dataset, the networks were mainly connected by members of the movement in the first-order zone, and news correspondents in second-order and tertiary zones. It was discovered that in all of the datasets, the #FreeJahar social movement appears to be the contributing factor that connects the disparate community into a small-world network by maintaining the ties via the strengthening of the interactions between the news events and the communities. Whilst the social movement acts to maintain the ties of the network, the activities within the cluster is fuelled by the news events as the natural reaction in active discussions is to utilise news items as accusations and for presenting arguments. This is confirmed by the Betweenness and Closeness centrality measures, which ranked highest within the cluster when the activities of the movement are at its peak in the dataset analysed here. In latter datasets, when the centralities of the social movement were disrupted by Twitter account suspensions, the movement begins to fragment and is eventually swallowed up by activities of the opposing group. As a result, the ranking of the centralities fluctuates and shifts between the #FreeJahar cluster and the major news channels. The social movement also confirms that, across the datasets, that an evenly distributed centrality has highly independent actors and high activities within the group.

The documentation and understanding of how small-world networks form and are strengthened and maintained via interactions in the real world is important. The activities of ad-hoc communities, online activists and social movements that have appeared in recent times have contributed to major changes to institutional reforms, transparency and the establishment of better policies for concerned communities. Other coordinated activities have affected the world in alternate ways. Identifying small-world activities could allow interested parties in assisting and strengthening certain communities for a good cause, and the same can be used for disrupting activities that are detrimental to society.

**Acknowledgements**

The author acknowledges the financial support from the International Doctoral Innovation Centre, Ningbo Education Bureau, Ningbo Science and Technology Bureau, China's MoST and The University of Nottingham. The project is partially supported by NBSTB Project 2012B10055

**Disclosure Statement**

No competing financial interests exist.